\documentclass[11pt]{article}
\usepackage{epsfig} 
\newcommand{\Dslash}{D \! \! \! \! /}

\newcommand{\half}{\mbox{\small{$\frac{1}{2}$}}}

\newcommand{\Nc}{N_{\!c}}
\newcommand{\Nf}{N_{\!f}}
\newcommand{\NF}{N_{\!F}}
\newcommand{\NA}{N_{\!A}}

\begin{document}
\title{Natural solution in the refined Gribov-Zwanziger theory}
\author{D.J. Thelan \& J.A. Gracey, \\ Theoretical Physics Division, \\ 
Department of Mathematical Sciences, \\ University of Liverpool, \\ P.O. Box 
147, \\ Liverpool, \\ L69 3BX, \\ United Kingdom.} 
\date{}
\maketitle 

\vspace{5cm} 
\noindent 
{\bf Abstract.} We analyse the one loop effective action of the
Gribov-Zwanziger Lagrangian and use the local composite operator formalism to
include the most general Becchi-Rouet-Stora-Tyutin (BRST) dimension two mass
operator for the localizing ghost fields. We show that the energetically
favourable colour channel corresponds to what is known as the ${\cal R}$
direction.

\vspace{-15cm}
\hspace{13cm}
{\bf LTH 1007}

\newpage 

In recent years there has been interest in understanding the infrared behaviour
of the gluon and Faddeev-Popov ghost propagators in Quantum Chromodynamics 
(QCD). This is motivated in the main by their relation to confinement. Gluons, 
which are the quanta mediating the strong nuclear force, are not observed in 
nature unlike the other vector bosons in the full standard model. In this 
respect the gluon propagator does not have the canonical behaviour associated 
with an observed fundamental particle which is the presence of a simple pole at
the mass shell value. While its high energy asymptotic properties are similar 
to those of say a photon, in the low energy regime a pole at zero momentum does
not apparently emerge. Evidence for this is provided by several approaches.
These are lattice gauge theory, Schwinger-Dyson methods and the Hamiltonian 
approach (in the Coulomb gauge), \cite{1}. The majority activity in recent 
years has primarily centred on the Landau gauge. Though in this gauge, which we 
concentrate on here, there are several scenarios.  In one case at zero momentum
the gluon propagator freezes to a non-zero finite value without any singular 
behaviour. A selection of articles demonstrating this are, for example, 
\cite{2,3,4,5,6,7,8,9,10,11}. However, in analysing non-abelian gauge theories 
at low energies one has to be aware of global considerations. In the case of 
Landau gauge fixed QCD the main problem one has to be aware of is the Gribov 
problem, \cite{12}. Stated briefly it is not possible to fix the gauge globally
as there are different gauge copies satisfying the same gauge fixing condition 
which have to be factored out of the path integral to ensure that there is no 
double counting. Although there are additional subtle issues as to whether 
accounting for such Gribov copies produces a unique gauge configuration, Gribov
demonstrated, \cite{12}, that their presence affected the properties of the 
gluon and Faddeev-Popov propagators at low momenta. The former is suppressed 
and freezes to a zero value while the latter enhances with a double pole 
structure in $p^2$ where $p$ is the momentum. Clearly this is not the 
behaviour found in recent years from lattice and some Schwinger-Dyson 
solutions, \cite{2,3,4,5,6,7,8,9,10,11}. On terminology the original Gribov 
scenario of \cite{12} is referred to as the scaling solution and the non-zero 
freezing is termed the decoupling solution, \cite{11}. Indeed recently various 
lattice analyses have shed some light as to why the latter emerges in the data
as opposed to the scaling case, \cite{13,14}. However, Schwinger-Dyson 
solutions do find the scaling solution too. Moreover, the Coulomb gauge 
Hamiltonian approach strongly supports the Gribov confinement picture and
connects the dual Meissner effect with the Gribov-Zwanziger confinement
analysis, \cite{1,15,16,17}.

After Gribov's definitive analysis of the gauge fixing problem in the Landau
gauge, \cite{12}, the non-local resultant Lagrangian was localized in a series 
of articles, \cite{18,19,20,21,22,23,24,25,26}. This produced a local 
renormalizable Lagrangian which meant that one could carry out loop 
computations. For instance, the two loop gap equation satisfied by the Gribov 
mass $\gamma$, which derives from the path integral cutoff, was determined in 
\cite{12,25,27}. Also the zero momentum freezing of the gluon propagator and 
the Faddeev-Popov ghost enhancement were confirmed at one and two loops 
respectively. However, the localization process of 
\cite{18,19,20,21,22,23,24,25,26} which Zwanziger constructed required two sets
of extra localizing spin-$1$ ghost fields. One set is bosonic and the other is 
Grassmann. These play a passive role at high energy but affect the low momentum
behaviour of the propagators. Indeed they have interesting dynamics in 
themselves. For instance, it was shown in \cite{28} that the adjoint projection
of the bosonic localizing ghost had an enhanced behaviour in its longitudinal 
part. However, like the original Gribov formulation, this pure Gribov-Zwanziger
Lagrangian does not produce the decoupling properties observed in current data.
To model this the Gribov-Zwanziger Lagrangian was refined in \cite{29} to
include a mass operator for the localizing ghost fields. Such a mass term 
alters the infrared properties of their propagators as well as that of the 
gluon. In particular the gluon has a non-zero value at zero momentum. The 
analysis was based on the local composite operator (LCO) formalism, 
\cite{30,31,32}, and constructed the effective potential satisfied by the mass 
operator. It was then shown that the potential had a minimum at a non-zero 
value indicating that the condensation of the operator would produce a mass 
term for the localizing ghosts thereby modelling the non-zero frozen gluon 
propagator behaviour, \cite{29}. However, the analysis of \cite{29} did not use
the most general possible localizing ghost mass operator. In \cite{33} the most
general dimension two BRST invariant operator was considered based on all the 
possible colour tensors. One feature was that the frozen gluon propagator 
behaviour did not have a unique solution from the LCO mechanism. Indeed it was
shown in \cite{33} that other colour structures not considered in \cite{29} 
could reproduce lattice data. Therefore to reconcile whether there is a
preferred colour tensor structure we will extend here the analysis of \cite{29}
to the general case of \cite{33} and complete the programme begun in \cite{33}.
The aim is to see if there is indeed a {\em unique} minimum solution to the 
corresponding LCO effective potential for the general operator. If there is a 
unique minimum we can therefore regard this as the energetically favourable 
operator condensation colour direction. Moreover, once determined the 
properties of the corresponding propagators at low momentum can therefore 
provide potential tests on future data to ascertain whether this is the 
underlying Lagrangian structure. 

To set the background to the problem we recall Gribov's observation, \cite{12},
that the QCD action has to be modified to account for the effect of copies 
deriving from the global gauge fixing ambiguity. The subsequent non-local 
Gribov Lagrangian which depends on the Gribov mass is, \cite{12,25},
\begin{equation}
L^{\mbox{\footnotesize{Gribov}}} ~=~ L^{\mbox{\footnotesize{QCD}}} ~+~
\frac{\gamma^4}{2} f^{eac} f^{ebd} A^a_\mu \left( \frac{1}{\partial^\nu D_\nu} 
\right)^{cd} A^{b\,\mu} ~-~ \frac{d \NA \gamma^4}{2g^2}
\label{lagrib}
\end{equation}
where $A^a_\mu$ is the gluon, $\NA$ is the adjoint representation dimension 
with $1$~$\leq$~$a$~$\leq$~$\NA$, $d$ is the spacetime dimension and
\begin{equation}
L^{\mbox{\footnotesize{QCD}}} ~=~ -~ \frac{1}{4} G_{\mu\nu}^a
G^{a \, \mu\nu} ~-~ \frac{1}{2\alpha} (\partial^\mu A^a_\mu)^2 ~-~
\bar{c}^a \partial^\mu D_\mu c^a ~+~ i \bar{\psi}^{iI} \Dslash \psi^{iI}
\end{equation}
is the QCD Lagrangian for $\Nf$ massless quarks $\psi^{iI}$ with
$1$~$\leq$~$i$~$\leq$~$\Nf$ and $1$~$\leq$~$I$~$\leq$~$\NF$ where $\NF$ is the
fundamental representation dimension and $c^a$ is the Faddeev-Popov ghost. The 
presence of $\gamma$ results from imposing the no pole condition, \cite{12}, 
which equates to defining the Gribov horizon in configuration space. From  
(\ref{lagrib}) this is, \cite{12,25},
\begin{equation}
f^{eac} f^{ebd} \left\langle A^a_\mu \left( \frac{1}{\partial^\nu D_\nu}
\right)^{cd} A^{b\,\mu} \right\rangle ~=~ \frac{d \NA}{g^2} ~.
\end{equation}
Given that the gluon propagator depends on $\gamma$ this condition constrains
$\gamma$ to satisfy a gap equation. In other words, \cite{12,20,25}, only when 
the gap equation is imposed is one in the gauge theory. To circumvent the
inability to perform computations with a Lagrangian with a non-local term, in a
series of articles, \cite{18,19,20,21,22,23,24,25,26}, Zwanziger localized the 
horizon term of (\ref{lagrib}) with the introduction of localizing ghost 
fields. In the current formulation of this these fields are $\xi_\mu^{ab}$, 
$\rho_\nu^{ab}$, $\omega_\mu^{ab}$ and $\bar{\omega}_\mu^{ab}$ where the first 
two are bosonic and the other two are Grassmann. The latter play the same role 
to the bosonic localizing ghosts as the Faddeev-Popov ghosts do for the gauge 
field. The full localized Lagrangian is, \cite{25},
\begin{eqnarray}
L^{\mbox{\footnotesize{GZ}}} &=& L^{\mbox{\footnotesize{QCD}}} ~+~
\frac{1}{2} \rho^{ab \, \mu} \partial^\nu \left( D_\nu \rho_\mu
\right)^{ab} ~+~ \frac{i}{2} \rho^{ab \, \mu} \partial^\nu
\left( D_\nu \xi_\mu \right)^{ab} ~-~ \frac{i}{2} \xi^{ab \, \mu}
\partial^\nu \left( D_\nu \rho_\mu \right)^{ab} \nonumber \\
&& +~ \frac{1}{2} \xi^{ab \, \mu} \partial^\nu \left( D_\nu \xi_\mu
\right)^{ab} ~-~ \bar{\omega}^{ab \, \mu} \partial^\nu \left( D_\nu \omega_\mu
\right)^{ab} ~-~ \frac{1}{\sqrt{2}} g f^{abc} \partial^\nu
\bar{\omega}^{ae}_\mu \left( D_\nu c \right)^b \rho^{ec \, \mu} \nonumber \\
&& -~ \frac{i}{\sqrt{2}} g f^{abc} \partial^\nu \bar{\omega}^{ae}_\mu
\left( D_\nu c \right)^b \xi^{ec \, \mu} ~-~ i \gamma^2 f^{abc} A^{a \, \mu}
\xi^{bc}_\mu ~-~ \frac{d \NA \gamma^4}{2g^2} ~.
\label{lagloc}
\end{eqnarray}
Its properties are well established, \cite{25,26,27}. For the purposes of this 
article the key ones are that the Faddeev-Popov ghosts as well as 
$\omega_\mu^{ab}$ enhance in the infrared when the gap equation for $\gamma$ 
is satisfied. Hence the Kugo-Ojima confinement criterion, \cite{34}, is 
fulfilled. Indeed more recently an analysis of this condition for other gauges 
such as Coulomb and the maximal abelian gauge has been provided in \cite{35}. 
There the Kugo-Ojima confinement condition was generalized to these additional 
gauges and criteria were given to differentiate between the Higgs or Coulomb
phases of BRST symmetric gauge theories and the colour confining phase. Indeed
this represents progress towards having a {\em universal} criterion for colour 
confinement. In addition there is an infrared enhancement for the bosonic 
localizing ghosts in (\ref{lagloc}). More specifically it has been shown in 
\cite{28} that the adjoint projection of the longitudinal part of the 
$\xi_\mu^{ab}$ propagator enhances in the infrared. It was argued that this 
feature reflected the Goldstone boson associated with the spontaneous breaking 
of a BRST related symmetry of the Lagrangian in the presence of a constraint. 
The final property of (\ref{lagrib}) which is relevant is that the gluon 
propagator is suppressed in the infrared. The last property can be seen in the 
propagators of (\ref{lagloc}) which are
\begin{eqnarray}
\langle A^a_\mu(p) A^b_\nu(-p) \rangle &=& -~
\frac{\delta^{ab}p^2}{[(p^2)^2+C_A\gamma^4]} P_{\mu\nu}(p) ~~~,~~~
\langle A^a_\mu(p) \xi^{bc}_\nu(-p) \rangle ~=~
\frac{i f^{abc}\gamma^2}{[(p^2)^2+C_A\gamma^4]} P_{\mu\nu}(p) \nonumber \\
\langle \xi^{ab}_\mu(p) \xi^{cd}_\nu(-p) \rangle &=& -~
\frac{\delta^{ac}\delta^{bd}}{p^2}\eta_{\mu\nu} ~+~
\frac{f^{abe}f^{cde}\gamma^4}{p^2[(p^2)^2+C_A\gamma^4]} P_{\mu\nu}(p) ~~,~~
\langle A^a_\mu(p) \rho^{bc}_\nu(-p) \rangle ~=~ 0 \nonumber \\
\langle \rho^{ab}_\mu(p) \rho^{cd}_\nu(-p) \rangle &=&
\langle \omega^{ab}_\mu(p) \bar{\omega}^{cd}_\nu(-p) \rangle ~=~ -~
\frac{\delta^{ac}\delta^{bd}}{p^2} \eta_{\mu\nu} ~~,~~
\langle \xi^{ab}_\mu(p) \rho^{cd}_\nu(-p) \rangle ~=~ 0 
\label{gzprop}
\end{eqnarray}
where $P_{\mu\nu}(p)$~$=$~$\eta_{\mu\nu}$~$-$~$p_\mu p_\nu/p^2$. The gluon 
suppression has been checked at one loop, \cite{36}, which is due in the main 
to the fact that (\ref{lagloc}) is renormalizable, \cite{25,37,38}, allowing 
one to perform loop computations. 

In summarizing these general features of the Gribov construction which persist
in loop calculations, it is evident that they are not observed on the lattice,
\cite{2,3,4,5,6,7,8,9,10,11}. Instead the numerical data indicates that the 
gluon propagator freezes to a non-zero value and the Faddeev-Popov ghost is not
enhanced in the infrared. This decoupling behaviour has been observed in one 
set of Schwinger-Dyson solutions, \cite{11}. However, in \cite{29} a BRST 
invariant mass operator for the localizing ghosts was introduced and its effect
on the structure of the theory in the infrared was analysed. Briefly a frozen 
gluon propagator and non-enhanced Faddeev-Popov ghost propagator emerged in the 
quantum analysis which mimics the lattice observations. It was subsequently 
pointed out in \cite{33} that the particular choice of BRST invariant mass 
operator of \cite{29} was {\em not} unique to modelling the decoupling 
solution. Indeed it was not the most general possible BRST dimension two 
operator from a group theoretic point of view. While one could in principle add
a mass operator for the localizing ghosts to (\ref{lagloc}) such a term would 
have no origin in the original Gribov Lagrangian, (\ref{lagrib}). Therefore, a 
mass term for the extra fields can be established via a non-zero vacuum 
expectation value for the mass operator. In other words if one computed the 
effective potential for the mass operator and found that it was a minimum at a 
non-zero value which was energetically more favourable than the massless 
localizing ghost case then the mass operator would condense. Thus the non-zero 
vacuum expectation value would produce the necessary masses to model a frozen 
gluon propagator and non-enhanced Faddeev-Popov ghost propagator. In \cite{29} 
such an analysis was performed for the operator considered there. However, as 
noted in \cite{33} by restricting the seed operator to a specific colour 
direction in colour space it was not clear whether the vacuum solution which 
emerged was the energetically most favourable one. Therefore, we provide the 
analysis for the most general BRST dimension two localizing ghost operator and 
aim to establish the direction in colour space which is the energetically 
favourable.

The most general BRST dimension two operator was introduced in \cite{33} and in
the same (non-orthogonal) basis is 
\begin{equation}
{\cal O} ~=~ \left[ \mu_{{\cal Q}}^2 \delta^{ac} \delta^{bd} ~+~
\mu_{{\cal W}}^2 f^{ace} f^{bde} ~+~ \frac{\mu_{{\cal R}}^2}{C_A}
f^{abe} f^{cde} ~+~ \mu_{{\cal S}}^2 d_A^{abcd} ~+~
\frac{\mu_{{\cal P}}^2}{\NA} \delta^{ab} \delta^{cd} ~+~ \mu_{{\cal T}}^2
\delta^{ad} \delta^{bc} \right] {\cal O}^{abcd}
\label{opdef}
\end{equation}
where the colour directions are each associated with a mass parameter, 
$\mu_{{\cal I}}^2$. We will use the label ${\cal I}$ to indicate the various 
possible colour channels defined by the colour tensor. For reference in 
\cite{29} channel ${\cal Q}$ was the main focus. In (\ref{opdef}) $f^{abc}$ are
the structure function of the colour group and, \cite{39}, 
\begin{equation}
d_A^{abcd} ~=~ \frac{1}{6} \mbox{Tr} \left( T_A^a T_A^{(b} T_A^c T_A^{d)}
\right)
\end{equation}
which is totally symmetric in its colour indices. The field content of the
operator is, \cite{33},  
\begin{equation}
{\cal O}^{abcd} ~=~ \frac{1}{2} \left[ \rho^{ab} \rho^{cd} ~+~
i \xi^{ab} \rho^{cd} ~-~ i \rho^{ab} \xi^{cd} ~+~ \xi^{ab} \xi^{cd} \right] ~-~
\bar{\omega}^{ab} \omega^{cd} 
\label{brstop}
\end{equation}
which is BRST invariant. Its anomalous dimension satisfies a Slavnov-Taylor 
identity and is simply related to the Faddeev-Popov ghost anomalous dimension,
\cite{29,33}.

To proceed to the effective potential that the operator satisfies we outline 
the local composite operator method, \cite{30,31,32}, in general terms. If we 
have an operator ${\cal O}$ then it is coupled to a source $J$ and the 
generating functional, $W[J]$, is constructed
\begin{equation}
e^{-W[J]} ~=~ \int {\cal D} \Phi_{\mbox{\footnotesize{o}}} \exp 
\left[ ~-~ S ~+~ \int d^d x J_{\mbox{\footnotesize{o}}}
{\cal O}_{\mbox{\footnotesize{o}}} ~+~ \frac{1}{2}
\zeta_{\mbox{\footnotesize{o}}} J_{\mbox{\footnotesize{o}}}^2 \right]
\label{Wo}
\end{equation}
where $\Phi$ represents the fields of the action $S$ and the subscript 
${}_{\mbox{\footnotesize{o}}}$ corresponds to a bare quantity. After 
renormalization one has, \cite{30,31,32}, 
\begin{equation}
e^{-W[J]} ~=~ \int {\cal D} \Phi \exp \left[ ~-~ S ~+~ Z_{{\cal O}} \int d^d x 
J {\cal O} ~+~ \frac{1}{2} \left( \zeta + \delta \zeta \right) J^2 \right] ~.
\end{equation}
The quantity $\zeta$ is known as the LCO parameter and is a non-perturbative
function of the coupling constant. It is defined to ensure that the 
renormalization group equation satisfied by $W[J]$ is homogeneous, 
\cite{30,31,32}. As it has similar properties to a coupling constant it 
undergoes renormalization but we use the same notation for the counterterm, 
$\delta \zeta$, as \cite{30,31,32}. The method to determine the explicit 
divergences contributing to $\delta \zeta$ was developed from the ideas of 
\cite{30} in \cite{40}. If one denotes the renormalization group function 
associated with the renormalization of $\zeta$ by $\delta(g)$ in the notation 
of \cite{30} then $\zeta(g)$ is defined by the solution of
\begin{equation}
\mu \frac{\partial \zeta}{\partial \mu} ~=~ 2 \gamma_{{\cal O}}(g) \zeta ~+~
\delta(g) 
\end{equation}
where $\gamma_{{\cal O}}(g)$ is the anomalous dimension of the operator whose
effective potential we are interested in. Once $\zeta(g)$ is defined to the 
loop order required one evaluates $W[J]$ or applies a Hubbard-Stratonovich 
transformation to translate the source in the exponential within the path 
integral definition to a linear dependence, \cite{30}. This allows one to 
construct the effective potential using the standard procedures. Here the 
structure of $W[J]$ will be sufficient for our purposes as its dependence on 
$\mu_{{\cal I}}^2$ will determine the energetically favourable colour 
direction. 

One main difference from the earlier application of the LCO formalism to
$\half A^a_\mu A^{a\,\mu}$ is that the operator (\ref{opdef}) can be regarded 
as a sum of different operators which are colour projections of (\ref{brstop}).
In this respect one should have a vector of sources, $J_{{\cal I}}$, so that 
the operator source seed term in the initial application of the LCO formalism 
in (\ref{Wo}) is 
\begin{equation}
\left[ J_{{\cal Q}} \delta^{ac} \delta^{bd} ~+~
J_{{\cal W}} f^{ace} f^{bde} ~+~ \frac{J_{{\cal R}}}{C_A}
f^{abe} f^{cde} ~+~ J_{{\cal S}} d_A^{abcd} ~+~
\frac{J_{{\cal P}}}{\NA} \delta^{ab} \delta^{cd} ~+~ J_{{\cal T}}
\delta^{ad} \delta^{bc} \right] {\cal O}^{abcd} ~.
\end{equation}
From this we have performed the summation of one loop leg graphs for the
$SU(\Nc)$ colour group and found
\begin{equation}
W[J] ~=~ -~ \frac{d\NA\gamma^4 Z_\gamma^4}{2g^2 Z_g^2} ~+~
\frac{d\NA\zeta J \gamma^2}{g^2} ~+~
\frac{(d-1)\NA}{2} \int_k \ln \left[ k^2 \left[ k^2
+ \frac{C_A\gamma^4}{[k^2+J]} \right] \right] ~+~ O(g^2)
\label{Wval}
\end{equation}
where $Z_\gamma$ and $Z_g$ are the respective renormalization constants for
$\gamma$ and $g$ and
\begin{equation}
J ~=~ \left[ J_{{\cal Q}} ~+~ J_{{\cal R}} ~-~ J_{{\cal T}} ~+~ 
\frac{C_A}{2} J_{{\cal W}} \right]
\end{equation} 
is the combination of sources which emerges from the formulation. The second
term of (\ref{Wval}) has no one loop divergences, \cite{29}. We recall that to 
proceed to the effective potential one introduces a field $\sigma(x)$ which is 
the field which couples linearly to the source, \cite{30,31,32}, and in effect 
corresponds to the original composite operator. Consequently the effective 
potential for $\sigma$ emerges from the constant field value of the effective 
action for the operator given by
\begin{equation}
\Gamma[\sigma] ~=~ W[J] ~-~ \int d^4 x \, J(x) \sigma(x) 
\end{equation}
after a Legendre transformation, \cite{30,31,32}. Hence, the one loop effective 
potential is
\begin{eqnarray}
V(M^2) &=& -~ \frac{2 \NA \gamma^4}{g^2} + \frac{4\NA\zeta M^2 \gamma^2}{g^2} 
\nonumber \\
&& +~ \frac{\NA}{64 \pi^2} \left[ 7 C_A \gamma^4 
+ M^4 \ln \left( \frac{M^2}{\mu^2} \right) 
+ 3 M^2 \sqrt{[M^4 - 4C_A \gamma^4]} \ln \left( \frac{a_+^2}{\mu^2} \right) 
\right. \nonumber \\
&& \left. 
-~ \frac{3}{4} [M^4 - 4 C_A \gamma^4] \ln \left( \frac{C_A \gamma^4}{\mu^4}
\right) 
- \frac{3}{2} M^2 \sqrt{[M^4 - 4C_A \gamma^4]} 
\ln \left( \frac{C_A \gamma^4}{\mu^4} \right) \right] 
\end{eqnarray}
where $a_+^2$~$=$~$\half \left[ M^2 + \sqrt{[M^4-4C_A\gamma^4]} \right]$ and
$\mu$ is the scale introduced to ensure that $g$ is dimensionless in 
$d$-dimensions. In the present context the effective potential depends on the 
{\em unique} combination of masses given by 
\begin{equation}
M^2 ~=~ \left[ \mu_{{\cal Q}}^2 ~+~ \mu_{{\cal R}}^2 ~-~ \mu_{{\cal T}}^2 ~+~
\frac{C_A}{2} \mu_{{\cal W}}^2 \right]
\label{coldir}
\end{equation}
where $M^2$ represents the constant field value of the corresponding $\sigma$
field. The potential has an absolute minimum at this value of $M^2$ as it is
this specific combination which minimizes the potential. Any deviation away 
from this combination will increase it. However, considering the overall 
situation this means that not only does the original operator ${\cal O}^{abcd}$
condense, it does so in a particular colour direction. To see what this is one 
computes the different combinations of $\mu_{{\cal I}}^2$ which emerge when the
colour tensor 
\begin{equation}
{\cal T}^{abcd} ~=~ \left[ \mu_{{\cal Q}}^2 \delta^{ac} \delta^{bd} ~+~
\mu_{{\cal W}}^2 f^{ace} f^{bde} ~+~ \frac{\mu_{{\cal R}}^2}{C_A}
f^{abe} f^{cde} ~+~ \mu_{{\cal S}}^2 d_A^{abcd} ~+~
\frac{\mu_{{\cal P}}^2}{\NA} \delta^{ab} \delta^{cd} ~+~ \mu_{{\cal T}}^2
\delta^{ad} \delta^{bc} \right] 
\end{equation}
is multiplied by each constituent tensor in turn. It is straightforward to see
that, for $SU(\Nc)$, ${\cal T}^{abcd} f^{abe} f^{cde}$~$=$~$C_A M^2$. Thus at 
one loop from the LCO effective potential the general operator condensation is 
in the ${\cal R}$ colour direction since
\begin{equation}
\langle {\cal O}^{abcd} \rangle ~\propto~ f^{abe} f^{cde} ~.
\label{Ovev}
\end{equation}
This preferred colour direction was suggested in \cite{33} based on the 
structure of the propagators (\ref{gzprop}). Integrating over the localizing 
ghost propagators in the combination defining the BRST invariant dimension two 
operator, the only non-zero contribution comes from the massive propagator term
in the $\xi^{ab}_\mu$ propagator. Note that this only applies above two 
dimensions as there are potential infrared divergences in strictly two 
dimensions. Indeed the zero momentum behaviour of the gluon in two dimensions 
is different from that in three and four dimensions. Lattice data, 
\cite{41,42}, shows the gluon propagator is suppressed and vanishes at zero 
momentum unlike the non-zero freezing above two dimensions. In this respect the
scaling solution appears to be preferred over the decoupling one. Aside from 
this caveat it is worth stressing that our result, (\ref{Ovev}), is derived in 
an explicit one loop analysis and the observation of \cite{33} should be 
regarded as a trivial consistency check. Concerning three dimensions it is also
the case that the ${\cal R}$ channel is the energetically favourable one. This 
can be seen from the summation of the graphs leading to $W[J]$ since the 
emergence of the combination (\ref{coldir}) for the mass dependence is due to 
the underlying group theory. Integration over the loop momentum does not affect
this. Instead it would lead to different values of the integral in (\ref{Wval})
when the loop integration is performed. If one accepts that the natural 
solution is the ${\cal R}$ colour direction then the next stage is the 
observation that the operator ${\cal O}^{abcd}$ condenses in this direction 
thereby giving non-zero masses to the localizing ghost fields. That one loop 
analysis was given in \cite{33}. The localizing ghost propagator corrections 
have properties which distinguish them from the ${\cal Q}$ solution and could 
be tested on the lattice.

To conclude we have extended the analysis of \cite{29} to consider the most 
general dimension two BRST invariant localizing ghost operator in the 
Gribov-Zwanziger Lagrangian of \cite{33}. The computation indicates that at 
least to one loop the energetically favoured colour condensation channel is the
${\cal R}$ one. To proceed one would have to continue to the next loop order. 
This would be a huge task given the multiscale nature of the two loop massive 
vacuum bubble graphs which would arise. However, in the interim one hope would 
be that the infrared structure of the localizing ghost propagators could be 
analysed by other techniques. While the lattice could provide numerical data 
the definition of an object on the lattice corresponding to say $\xi^{ab}_\mu$ 
is not straightforward. So a Schwinger-Dyson approach may offer the best avenue
for an independent analysis.

\vspace{1cm}
\noindent
{\bf Acknowledgements.} This work was carried out in part with the support of 
an STFC studentship (DJT).

\end{document}